\documentclass[twocolumn,showpacs]{revtex4}

\usepackage[latin1]{inputenc}
\usepackage{graphicx}%
\usepackage{dcolumn}
\usepackage{amsmath}

\makeatletter
\def\btt#1{\texttt{\@backslashchar#1}}%
\DeclareRobustCommand\bblash{\btt{\@backslashchar}}%
\makeatother

\begin{document}


 \title{Physics of Autonomous Driving based on Three-Phase Traffic Theory}

\mark{Autonomous Driving based on Three-Phase Traffic Theory}

\author{   
Boris S. Kerner$^1$}

 \affiliation{$^1$
Physics of Transport and Traffic, University Duisburg-Essen,
47048 Duisburg, Germany}



\begin{abstract} 
We have revealed physical features of autonomous driving in the framework of the three-phase traffic theory
  for which
there is {\it no}  fixed  time headway to the preceding vehicle. A comparison with
 the  classical model approach to autonomous driving 
	for which an autonomous driving vehicle tries to reach a fixed
	(desired or $\lq\lq$optimal'') time headway to the preceding vehicle
has been made. It turns out that  
autonomous driving in the framework of the three-phase traffic theory exhibits
 the following advantages in comparison with the classical model of 
autonomous driving:
(i) The absence of  string instability. (ii) 
Considerably smaller speed disturbances at road bottlenecks.
(iii) Autonomous driving vehicles based on the three-phase theory decrease
the probability of traffic breakdown  at the bottleneck in
  mixed
  traffic flow consisting of human driving and autonomous driving vehicles; on the contrary, even a single   autonomous driving
vehicle based on the classical approach can   provoke traffic breakdown at the bottleneck in   mixed traffic flow.
\end{abstract}

\maketitle

\begin{figure}
\begin{center}
\includegraphics*[width=8 cm]{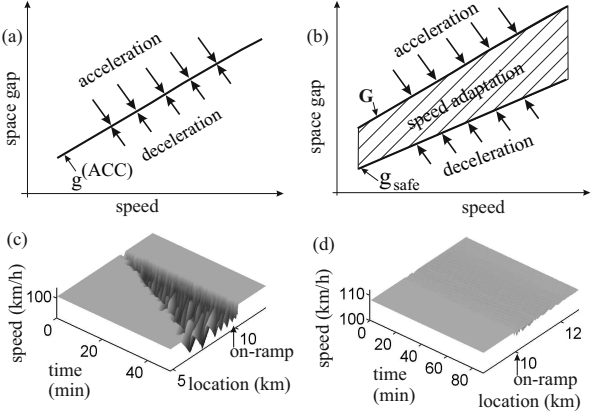}
\end{center}
\caption[]{String instability of the classical ACC model (\ref{ACC_General})  (see a version of ACC model with discrete time 
  used for simulations in~\cite{Kerner_Review3_Aut})
(a, c) and string stability of  TPACC model (\ref{TPACC_main})--(\ref{next2_TPACC})
(b, d) in traffic flow consisting of 100$\%$
of autonomous driving vehicles on a single-lane road with on-ramp bottleneck located at $x=$10 km: 
(a) Qualitative speed dependence of desired space gap $g^{\rm (ACC)}=v\tau^{\rm (ACC)}_{\rm d}$ 
of ACC for which at a given speed there is a single operating point
 $g=g^{\rm (ACC)}$ (see, e.g.,~\cite{Levine1966A_Aut,Liang1999A_Aut,Liang2000A_Aut,Rajamani2012A_Aut,Swaroop1996A_Aut,Swaroop2001A_Aut,fail_Davis2004B9,fail_Davis2014C}). (b) Qualitative presentation of a
  part of 2D-region for operating points of TPACC in
the space-gap--speed plane:   at a given speed of TPACC-vehicle there are the infinity of 
operating points of TPACC~\cite{KernerPat20039_Aut3}.
(c, d) Simulations of string instability of ACC
at on-ramp bottleneck with the classical ACC model (\ref{ACC_General})
   with discrete time~\cite{Kerner_Review3_Aut}  (c)
and string  stability of TPACC
at on-ramp bottleneck with the TPACC model (\ref{TPACC_main})--(\ref{next2_TPACC}) (d):
 Speed in space and time.
Simulation parameters of ACC (c) and TPACC (d) are identical ones:
the on-ramp inflow rate $q_{\rm on}=320$ vehicles/h and the flow rate upstream of the bottleneck
$q_{\rm in}=2002.6$ vehicles/h, $\tau^{\rm (ACC)}_{\rm d}=\tau_{\rm p}=$1.3 s,
 $\tau_{\rm G}=$1.4 s, $K_{1}=0.3 \ s^{-2}$ and $K_{2}=K_{\rm \Delta v}=0.3 \ s^{-1}$;
   $a_{\rm max}=b_{\rm max}=3 \ {\rm m}/{\rm s}^2$, $v_{\rm free}=30$m/s (108 km/h),
vehicle length $d=$7.5 m.
In accordance with the  
desired time headway
$\tau^{\rm (ACC)}_{\rm d}=$1.3 s of the ACC-vehicles, the sum flow rate
$q_{\rm sum}=q_{\rm in}+q_{\rm on}=2322.6$ vehicles/h is related to  time headway
1.3 s between  vehicles in free flow.
}
\label{Eco_ACC}
\end{figure}

It is generally assumed that    future  vehicular traffic  
 is a mixed
  traffic flow consisting of human driving and autonomous driving vehicles~\cite{ACC_v,Levine1966A_Aut,Liang1999A_Aut,Liang2000A_Aut,Rajamani2012A_Aut,Swaroop1996A_Aut,Swaroop2001A_Aut,fail_Davis2004B9,fail_Davis2014C}.
  Autonomous driving vehicles should considerably
enhance   capacity of a traffic network that
 is limited by   traffic breakdown at network bottlenecks.
 For  an analysis of the dynamics of autonomous driving vehicles and its effect on
 traffic flow, we consider a simple case of 
	vehicular traffic on a single-lane road with an on-ramp bottleneck.
 On the single-lane road, no  vehicles can pass.
  For this reason, autonomous driving can be achieved through the use of 
	an adaptive cruise control (ACC) in a vehicle:
   An ACC-vehicle follows the preceding vehicle
	(that can be either
	a human driving vehicle or an  ACC-vehicle) automatically based on some ACC dynamics rules of motion.  
 In a classical ACC model, acceleration (deceleration) $a^{\rm (ACC)}$ of
the ACC vehicle is  determined  by   the space gap to the preceding vehicle $g$ 
and the relative speed $\Delta v=v_{\ell}-v$ measured by the ACC vehicle
 as well as by
  a desired time headway
$\tau^{\rm (ACC)}_{\rm d}$ of the ACC-vehicle to the 
 preceding 
vehicle  (Fig.~\ref{Eco_ACC} (a)) (see, e.g.,~\cite{Levine1966A_Aut,Liang1999A_Aut,Liang2000A_Aut,Rajamani2012A_Aut,Swaroop1996A_Aut,Swaroop2001A_Aut,fail_Davis2004B9,fail_Davis2014C} and references there): 
\begin{equation}
a^{\rm (ACC)} = K_{1}(g-v\tau^{\rm (ACC)}_{\rm d})+K_{2} (v_{\ell}-v),
 \label{ACC_General}
 \end{equation} 
where $v$ is the speed of the ACC-vehicle, $v_{\ell}$ is the speed of the preceding vehicle; here and below
  $v$, $v_{\ell}$, and $g$ are time-functions;
  $K_{1}$
and $K_{2}$ are   coefficients of  
   ACC adaptation.  
 It is well-known that there can be 
   string instability of a long enough platoon of ACC-vehicles 
	(\ref{ACC_General})~\cite{Levine1966A_Aut,Liang1999A_Aut,Liang2000A_Aut,Rajamani2012A_Aut,Swaroop1996A_Aut,Swaroop2001A_Aut,fail_Davis2004B9,fail_Davis2014C} that occurs under condition
	$K_{2}<(2-K_{1}(\tau^{\rm (ACC)}_{\rm d})^{2})/2\tau^{\rm (ACC)}_{\rm d}$ found by
	Liang and Peng~\cite{Liang1999A_Aut}.

However,  
 there is another
 basic physical problem   of
the classical autonomous driving:  
 Even when  
  the  above-mentioned condition for string instability is not satisfied, i.e., any
 platoon of ACC-vehicles is stable, already a small share of ACC-vehicles in mixed traffic flow can deteriorate
traffic  while provoking traffic breakdown at network bottlenecks~\cite{Kerner_Review3_Aut}.

A study of real field traffic data shows~\cite{Kerner1998E_Aut3} that
  the
basic feature of    classical autonomous driving systems -- a   desired time headway     
is fundamentally inconsistent with a basic behavior of
real drivers.      
To explain the empirical data, in the three-phase traffic theory is assumed
when a driver approaches a slower moving preceding vehicle and the driver cannot pass it, 
the driver  decelerates
within a synchronization space gap $G$ adapting 
the speed to the speed of the preceding vehicle   without caring what the precise  space gap $g$
 to the preceding vehicle is as long as it is not smaller than a safe space gap $g_{\rm safe}$~\cite{Kerner1998E_Aut3}.  
This speed adaptation occurs within the synchronization space gap that limits   
  a 2D-region of  synchronized flow states  (dashed region in Fig.~\ref{Eco_ACC} (b)) determined by
  conditions
	\begin{equation}
g_{\rm safe} \leq g \leq G.
\label{G_g_g_s}
\end{equation} 
	In other words, accordingly to (\ref{G_g_g_s}),
    drivers  
		do not try to reach a particular (desired or optimal)  time headway
to the preceding vehicle, but adapt the speed while keeping  time headway $\tau^{\rm (net)}=g/v$
in a range $\tau_{\rm safe} \leq \tau^{\rm (net)}\leq \tau_{\rm G}$, where
  $\tau_{\rm G}=G/v$, $\tau_{\rm G}$ is a synchronization time headway,
  $\tau_{\rm safe}=g_{\rm safe}/v$ is  a safe time headway 
	and it is assumed that the speed $v>0$.
In   inventions~\cite{KernerPat20039_Aut3}, 
we have assumed that to satisfy these empirical features of real traffic,
  acceleration (deceleration) of autonomous driving based on the three-phase theory
	(for short, {\bf t}hree-traffic-{\bf p}hase ACC -- TPACC)
  should be given by formula~\cite{KernerPat20039_Aut3} 
\begin{eqnarray}
\label{TPACC_Eq1}
a^{\rm (TPACC)}=K_{\rm \Delta v}(v_{\ell}-v) \quad 
 \textrm{at $g_{\rm safe} \leq g \leq G$}  
\end{eqnarray}
where  
$K_{\rm \Delta v}$ is a dynamic coefficient   ($K_{\rm \Delta v}>0$).

However,   no studies of autonomous driving based on the three-phase theory have been done up to now, and,
therefore,   the physics of   TPACC    has not   been known.
In this Letter, we reveal the physical features of TPACC and the effect of TPACC-vehicles on mixed traffic flow
as well as compare TPACC with     autonomous driving based on the classical
approach.

To understand physical features of autonomous driving in the framework of the three-phase theory,
we  introduce   the following TPACC
    model:
\begin{equation}
a^{\rm (TPACC)}=  
  \left\{
\begin{array}{ll}
K_{\rm \Delta v}(v_{\ell}-v) &  \textrm{at $g \leq G$} \\ 
K_{1}(g-v\tau_{\rm p})+K_{2} (v_{\ell}-v) &  \textrm{at $g> G$}, \\ 
\end{array} \right.
\label{TPACC_main5}  
 \end{equation}
where   $\tau_{\rm p}$ is a model parameter and it is assumed that $g\geq g_{\rm safe}$.
 All simulations 
of human driving vehicles in   mixed traffic flow   
 are made below with the Kerner-Klenov microscopic 
stochastic   
model with   discrete time $t=n\tau$, where $n=0,1,2,...$; $\tau=$1 s is   time step~\cite{KKl}. 
For this reason, we simulate TPACC-model (\ref{TPACC_main5})
with    discrete time $t=n\tau$. 
Respectively, TPACC-model (\ref{TPACC_main5}) should be rewritten as follows:
\begin{equation}
a^{\rm (TPACC)}_{n}=  
  \left\{
\begin{array}{ll}
K_{\rm \Delta v}(v_{\ell, n}-v_{n}) &  \textrm{at $g_{n} \leq G_{ n}$} \\ 
K_{1}(g_{n}-v_{n}\tau_{\rm p})+K_{2} (v_{\ell,n}-v_{n}) &  \textrm{at $g_{n}> G_{ n}$}, \\ 
\end{array} \right.
\label{TPACC_main}  
 \end{equation}
where  $G_{ n}=v_{n}\tau_{\rm G}$. 
When $g_{n} < g_{\rm safe, n}$, 
the TPACC-vehicle should move in accordance with some safety conditions
 to avoid collisions between vehicles (Fig.~\ref{Eco_ACC} (b)). A collision-free  
TPACC-vehicle motion is  described as made in~\cite{Kerner_Review3_Aut} 
for the classical model of ACC:  
\begin{equation}
\label{next1_TPACC}
 v^{\rm (TPACC)}_{{\rm c}, n} = v_{n}+\tau \max(-b_{\rm max},
\min(\lfloor a^{\rm (TPACC)}_{n} \rfloor, a_{\rm max})),  
\end{equation}
\begin{equation}
 v_{n+1} = \max(0, \min({v_{\rm free}, v^{\rm (TPACC)}_{{\rm c},n}, v_{{\rm s},n}})),
\label{next2_TPACC}
\end{equation}
 where $\lfloor z \rfloor$ denotes the 
integer part  of $z$ (note that Eqs.~(\ref{next1_TPACC}), (\ref{next2_TPACC}) for TPACC
are the same as those for the classical ACC  model with discrete time 
 considered in~\cite{Kerner_Review3_Aut});  the TPACC  acceleration and deceleration 
  are limited by   $a_{\rm max}$ and   
$b_{\rm max}$, respectively;
the speed   $v_{n+1}$  (\ref{next2_TPACC})
 at   time step $n+1$ is limited by the 
maximum speed   $v_{\rm free}$  and by the safe speed
 $v_{{\rm s},n}$ that have been chosen, respectively, the same as those in the
  model of human driving vehicles. 
The models of   human driving vehicles~\cite{KKl},  a discrete version
of the classical ACC model~\cite{Kerner_Review3_Aut}, the model of the
on-ramp bottleneck
as well as model parameters used in simulations   
have been reviewed  in Appendix~A of the book~\cite{Kerner_Book3}.
  
In accordance with Eq.~(\ref{next2_TPACC}),  condition
		$v^{\rm (TPACC)}_{{\rm c},n} \leq v_{{\rm s},n}$ is equivalent to condition
	$g_{n}\geq g_{\rm safe, n}$.  
		Under this condition,  
  from (\ref{TPACC_main})--(\ref{next2_TPACC}) it follows
 that when      time headway $\tau^{\rm (net)}_{n}=g_{n}/v_{n}$
 of the TPACC-vehicle   to the preceding vehicle is within the range
\begin{eqnarray}
\tau_{{\rm safe},n}\leq \tau^{\rm (net)}_{n}\leq \tau_{\rm G}, 
\label{TPACC_main_range} 
 \end{eqnarray}
the acceleration (deceleration) of the TPACC-vehicle does not depend on
  time headway, where $\tau_{{\rm safe},n}=g_{{\rm safe}, n}/v_{n}$ is a safe time headway
	and it is assumed that $v{_n}>0$.
Thus, in contrast with 
  the classical ACC model (\ref{ACC_General}) (Fig.~\ref{Eco_ACC} (a)),
 there is {\it no} 
fixed 	desired time headway to the preceding vehicle for the autonomous driving vehicle
based on the three-phase theory   
  (Fig.~\ref{Eco_ACC} (b)).

The speed adaptation  within the 2D-traffic flow states 
of the three-phase theory used in TPACC model (\ref{TPACC_main})--(\ref{next2_TPACC}) 
changes the dynamic behavior of autonomous 
driving   basically in comparison with the classical ACC model (\ref{ACC_General}).
 
 Simulations of   string instability of ACC-vehicles are shown 
in Fig.~\ref{Eco_ACC} (c). Speed disturbances in traffic flow 
consisting of 100$\%$ ACC-vehicles occur
 at an on-ramp bottleneck at which on-ramp inflow with the rate
$q_{\rm on}$ and upstream flow with the rate $q_{\rm in}$ merge.  
String instability of ACC-vehicles
   leads to the emergence of moving jams upstream of the bottleneck
	(Fig.~\ref{Eco_ACC} (c)). Contrarily, at   the  same set of the flow rates
	$q_{\rm on}$ and   $q_{\rm in}$
	as well as the same other model parameters  
	{\it no} string instability of any platoon of the TPACC-vehicles is realized:
In Fig.~\ref{Eco_ACC} (d), all 
speed disturbances occurring at the bottleneck decay
upstream of the bottleneck. It turns out that as long as  
   time headway   between TPACC-vehicle is within
 the range (\ref{TPACC_main_range}), speed disturbances decay over time.
This is because within this range the acceleration (deceleration) of
TPACC-vehicle does not depend on time headway.

\begin{figure}
\begin{center}
\includegraphics*[width=8 cm]{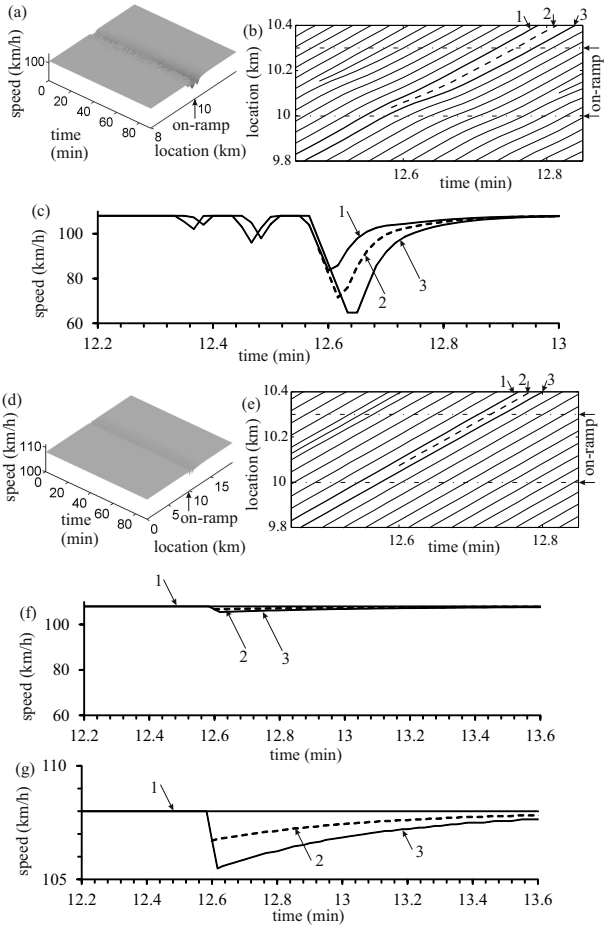}
\end{center}
\caption[]{Speed disturbances at on-ramp bottleneck in stable free flow
with 100$\%$ autonomous driving vehicles  for
 classical ACC model (\ref{ACC_General}) (see a version of ACC model with discrete time 
  used for simulations in~\cite{Kerner_Review3_Aut})
   (a--c) and for TPACC model
	(\ref{TPACC_main})--(\ref{next2_TPACC})
 (d--g):
(a, d) Speed in space and time.   (b, e) Fragments of vehicle trajectories; (b) is related to (a)
and   (e) is related to (d). (c, f, g) Microscopic speeds along vehicle trajectories
shown by, respectively, the same numbers in (b, e); (c) is related to (b)
and   (f, g) are related to (e); (f) and (g) show the same speed in different speed-scales.
Simulation parameters of ACC (a--c) and TPACC (d--g) are identical ones.
$K_{2}=K_{\rm \Delta v}=0.6 \ s^{-1}$. Other model parameters are the same as those in Fig.~\ref{Eco_ACC}.
}
\label{Fluc_ACC}
\end{figure}

\begin{figure}
\begin{center}
\includegraphics*[width=8 cm]{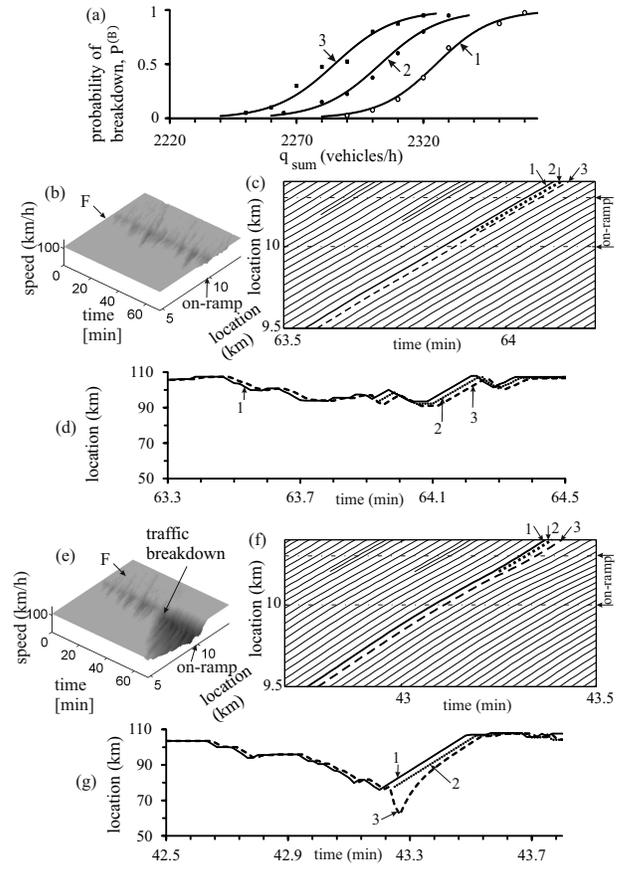}
\end{center}
\caption[]{Effect of a single autonomous driving vehicle on the probability of traffic breakdown
in mixed traffic flow    with only 2$\%$ of autonomous driving vehicles:
(a)  Probability of traffic breakdown at on-ramp bottleneck
as a function of the flow rate $q_{\rm sum}=q_{\rm in}+q_{\rm on}$
at a given flow rate $q_{\rm in}=$ 2000 vehicles/h; curve 1 is related to
traffic flow without autonomous driving vehicles as well as to
mixed traffic
 flow with TPACC-vehicles; curves 2 and 3 are related to mixed traffic flow with ACC-vehicles, respectively,
with $\tau^{\rm (ACC)}_{\rm d}=$1.3 s and 1.6 s. 
(b--g)
	Speed disturbances occurring at on-ramp bottleneck through 
a single TPACC-vehicle (b--d)
and a single ACC-vehicle (e--g):
(b, e)  Speed in space and time; F -- free flow.   (c, f) Fragments of vehicle trajectories.
 (d, g) Microscopic speeds along vehicle trajectories
shown by  the same numbers in (c, f), respectively. In (c, d, f, g),
vehicles 1 and 2 are manual driving vehicles whereas    vehicle 3 is ACC-vehicle in (c, d)
and  TPACC-vehicle in (f, g). In (b--g), $q_{\rm in}=$ 2000 vehicles/h, $q_{\rm on}=$ 280 vehicles/h;
other
  model parameters for ACC-vehicles and TPACC-vehicles are, respectively,
	the same as those in Fig.~\ref{Fluc_ACC}.
}
\label{Break_ACC}
\end{figure}

\begin{figure}
\begin{center}
\includegraphics*[width=8 cm]{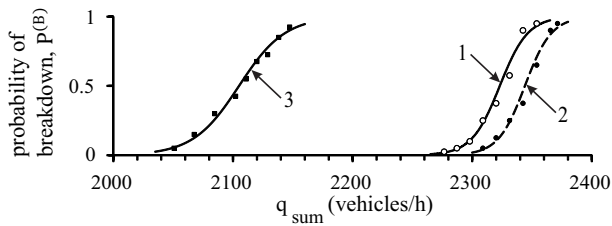}
\end{center}
\caption[]{Probability of traffic breakdown at on-ramp bottleneck
as a function of the flow rate $q_{\rm sum}=q_{\rm in}+q_{\rm on}$
at a given on-ramp inflow rate $q_{\rm on}=$ 320 vehicles/h  
   in mixed traffic flow
with 20$\%$ autonomous driving vehicles:
Curve 1 is related to
traffic flow without autonomous driving vehicles. Curves 2 and 3 are related to mixed traffic flow with
TPACC-vehicles (curve 2) and ACC-vehicles (curve 3). 
Simulation parameters of ACC   and TPACC   
   are, respectively, the same as those in Fig.~\ref{Fluc_ACC}.
}
\label{Break2_ACC}
\end{figure}

At a larger value of $K_{2}$ in (\ref{ACC_General})
as well as at the same 
desired time headway $\tau^{\rm (ACC)}_{\rm d}=$1.3 s and the same set of
the flow rates $q_{\rm on}$ and   $q_{\rm in}$
as those in Fig.~\ref{Eco_ACC},
	 platoons of   ACC-vehicles become   stable
	 (Fig.~\ref{Fluc_ACC} (a)).
However, it turns out that    considerable speed disturbances appear
at the bottleneck. This case is shown in Fig.~\ref{Fluc_ACC} (b, c) in which
  ACC-vehicle 2   merges from the on-ramp onto the main road following
  ACC-vehicle 1   moving on the main road.
To satisfy the
 desired time headway $\tau^{\rm (ACC)}_{\rm d}$, ACC-vehicle 2
should decelerates to a lower speed than the minimum speed of ACC-vehicle 1. 
This deceleration of ACC-vehicle 2 forces the following ACC-vehicle 3
to decelerate   while
approaching ACC-vehicle 2. Simulations show that
the occurrence of large speed disturbances at the bottleneck is a 
 basic problem of ACC-vehicles based on the classical
approach   in which the ACC-vehicles  try to reach  a
 desired time headway $\tau^{\rm (ACC)}_{\rm d}$.

 These large speed
disturbances at the bottleneck caused by ACC-vehicles
(Fig.~\ref{Fluc_ACC} (b, c)) do  not occur
in  traffic flow consisting of TPACC-vehicles
  (Fig.~\ref{Fluc_ACC} (d--g)).
	This is because within the range of time headway 
	(\ref{TPACC_main_range}) the acceleration (deceleration) of
an TPACC-vehicle does not depend on time headway. This explains
 small amplitudes of speed disturbances caused by TPACC-vehicles
  2  and 3 at the bottleneck (Fig.~\ref{Fluc_ACC} (e, f, g)).
The speed disturbances caused by TPACC-vehicles are so small (Fig.~\ref{Fluc_ACC} (f)) that
in the same speed-scale as that used for the classical ACC (Fig.~\ref{Fluc_ACC} (c)) they cannot
almost be resolved. Only at considerably larger speed-scale the speed disturbances become visible
(Fig.~\ref{Fluc_ACC} (g)).

Traffic breakdown in a flow of manual driving vehicles
is a phase transition from free flow (F) to synchronized flow (S) (F$\rightarrow$S transition)
  occurring in a metastable free flow with respect to the F$\rightarrow$S transition.
The larger the amplitude of
speed disturbances at the bottleneck, the more probable the nucleus occurrence
for the breakdown, i.e, the larger 
 the probability of traffic breakdown $P^{\rm (B)}$
at the bottleneck~\cite{Kerner_Review3_Aut,Kerner1998E_Aut3}. 

In the next future, we could expect
  a mixed traffic flow in which the share of autonomous driving vehicles is small
  (Fig.~\ref{Break_ACC}). Single  
	TPACC-vehicles moving in a such mixed traffic flow
	cause very small speed disturbances at the bottleneck   (Fig.~\ref{Break_ACC} (b--d)). 
	Indeed, we have found that probability of traffic breakdown remains
	in this mixed flow the same as that in traffic flow consisting of manual drivers only
	(curve 1 in Fig.~\ref{Break_ACC} (a)). Contrarily, probability of traffic breakdown
	can increase even when a very small amount of classical ACC-vehicles is in a
	mixed traffic flow
	(curves 2 and 3  in Fig.~\ref{Break_ACC} (a)). This deterioration of traffic through classical autonomous driving is explained by
  the occurrence of a large amplitude speed disturbance caused by  
a classical ACC-vehicle at the bottleneck (Fig.~\ref{Break_ACC} (e--g)): Already a single  ACC-vehicle
can initiate traffic breakdown at the bottleneck   (Fig.~\ref{Break_ACC} (e--g)).  

 If the share of autonomous driving vehicles in mixed traffic flow increases 
(Fig.~\ref{Break2_ACC}), the probability of traffic breakdown caused by ACC-vehicles
that deteriorate traffic 
can increase   considerably 
(Fig.~\ref{Break2_ACC}, curve 3). Contrarily, long enough platoons of TPACC-vehicles in mixed traffic flow
  decrease the breakdown probability (Fig.~\ref{Break2_ACC}, curve 2). 
This physical feature
of TPACC-vehicles is also explained by the speed adaptation effect of the three-phase theory that is the basis of TPACC  (\ref{TPACC_main5}):
 At each vehicle speed, the TPACC-vehicle   makes an arbitrary choice 
in   time headway that satisfies conditions (\ref{TPACC_main_range}). In other words,  
 the TPACC-vehicle 
accepts different values of time headway  at different times and does not control
  a fixed time headway to the preceding vehicle.

By autonomous driving in the framework of the three-phase theory
(TPACC)
there is {\it no} fixed desired time headway to the preceding vehicle.
In the Letter, we have shown that 
 this physical feature of 
TPACC leads to the following advantages in comparison with the classical approach to 
autonomous driving:
(i) The absence of  string instability. (ii) 
Considerably smaller speed disturbances at road bottlenecks.
(iii) Autonomous driving vehicles based on the three-phase theory decrease
the probability of traffic breakdown  at the bottleneck in   mixed traffic flow; on the contrary, even a single   autonomous driving
vehicle based on the classical approach can   provoke traffic breakdown at the bottleneck
in   mixed traffic flow.

I would like to thank Sergey Klenov for help and useful suggestions.
We thank our partners for their support in the project
   $\lq\lq$MEC-View -- Object detection for automated driving based on Mobile Edge Computing",
    funded by the German Federal Ministry of Economic Affairs and Energy.


\begin{thebibliography}{8.}
\bibitem{ACC_v}
Autonomous 
 driving vehicles are also called automated driving or  automatic driving or else self-driving vehicles. 
 Under automated driving or  automatic driving vehicles are often understood
autonomous 
 driving vehicles that are connected each other (vehicle-to-vehicle (V2V) communication)
or/and with infrastructure (V2X communication).
\end{thebibliography}
\end{document}